\documentclass[12pt]{article}
\usepackage[T1]{fontenc}
\usepackage[latin1]{inputenc}
\usepackage{geometry}
\usepackage{setspace}
\makeatletter

\providecommand{\LyX}{L\kern-.1667em\lower.25em\hbox{Y}\kern-.125emX\@}

 \newcommand{\lyxaddress}[1]{
   \par {\raggedright #1 
   \vspace{1.4em}
   \noindent\par}
 }

\makeatother

\begin{document}

\title{Additional considerations in the definition and renormalization of non-covariant
gauges}

\author{{\normalsize Satish D. Joglekar}\thanks{
email address: \emph{sdj@iitk.ac.in}
}\normalsize }

\maketitle

\lyxaddress{Department of Physics, I.I.T. Kanpur, Kanpur 208016 {[}INDIA{]}}

\begin{abstract}
{\normalsize In this work, we pursue further consequences of a general formalism
for non-covariant gauges developed in an earlier work (hep-th/0205042). We carry
out further analysis of the additional restrictions on renormalizations noted
in that work. We use the example of the axial gauge \( A_{3}=0 \). We find
that if multiplicative renormalization together with ghost-decoupling is to
hold, the {}``prescription-term{}'' (that defines a prescription) cannot be
chosen arbitrarily but has to satisfy certain non-trivial conditions (over and
above those implied by the validity of power counting) arising from the WT identitites
associated with the residual gauge invariance. We also give a restricted class
of solutions to these conditions. }{\normalsize \par}
\end{abstract}
The Yang-Mills theory in gauges other than the Lorentz gauges have been a subject
of wide research \cite{bns, l, 1}. These gauges have been used in a variety
of Standard Model calculations and in formal arguments in gauge theories \cite{bns,l}
(as well as in string theories). As compared to the covariant gauges, these
gauges have, however, not been fully developed \cite{dev}. Recently, an approach
that gives the \emph{definition} of non-covariant gauges in a Lagrangian path-integral
formulation, which moreover is compatible with the Lorentz gauges by construction,
has been given \cite{jm00} and exploited \cite{jm99} in various contexts such
as those related to the axial, planar and the Coulomb gauges. A general path-integral
framework, suggested by these results, that attempts to treat all these gauges
formally but \emph{rigorously} and hopefully \emph{completely} ( i.e. including
the treatment of all their problems) was presented recently \cite{j02}. Several
new observations regarding these gauges were recently made from such a framework
by simple and direct considerations \cite{j02}. This work presents further
results regarding the nature of the renormalization in axial gauges based on
this formulation and the results in \cite{j02}.

It was suggested in \cite{j02} that many of the ways of defining non-covariant
gauges including the one based on ref. \cite{jm00} can be formulated as a special
case of the path-integral\footnote{%
In the following, we use \( \phi  \) to generically denote all fields. 
} \begin{equation}
\label{ii}
W[J,K,\overline{K};\xi ,\overline{\xi }]=\int D\phi \; exp\{iS_{eff}[A,c,\overline{c},\psi ]+\varepsilon O[\phi ]+source-terms\}
\end{equation}
obtained by including \emph{an} \( \epsilon  \)-term\footnote{%
We may often require an \( \epsilon  \)-term of the form \( \varepsilon \int d^{4}xO[A,c,\overline{c};\varepsilon ] \);
i.e. with an \( \epsilon  \)-dependent \( O \).
}. We recognize that in order that (\ref{ii}) is mathematically well-defined,
this \( \epsilon  \)-term must, in particular, break the residual gauge invariance\footnote{%
A definition of the generalized residual gauge-invariance in the BRS-space has
been given in \cite{j02}.
} completely. In addition, to keep the discussion general enough and to cover
many of the ways suggested for dealing with these gauges, we do not necessarily
limit \( \epsilon  \) to have dimension two in the following, nor do we restrict
\( O \) to have local nature\footnote{%
We do not however imply that \emph{any} such \( \epsilon  \)-term will necessarily
be appropriate to define a gauge theory \emph{compatible with} the Lorentz gauges.
Existence (and construction) of an \emph{}\( \epsilon  \)-term which will serve
\emph{this purpose} is already known however. See e.g. \cite{jm00} and 5\( ^{th} \)
of ref. \cite{jm99}. 
}. 

We note that the various prescriptions, say the Leibbrandt-Mandelstam (L-M)
prescription for the light-cone gauges and the CPV for axial gauges etc, can
be understood\footnote{%
We however note some of the complications in the interpretation of \emph{double}
poles in CPV. See e.g. references \cite{bns,l}.
} as special cases of (\ref{ii}) {[}with rather complicated nonlocal \( O \){]}
and thus the following discussion should include these as special cases (For
more details, see ref. \cite{prep}). Generally, the axial poles are treated
by giving a way of interpreting the poles. They amount to replacing the naive
propagator (with \( \lambda \rightarrow 0 \))\begin{equation}
\label{nai}
\frac{-i}{k^{2}}\left[ g_{\mu \nu }-\frac{k_{\mu }\eta _{\nu} +k_{\nu }\eta _{\mu }}{\eta .k}+\frac{(\eta ^{2}+\lambda k^{2})k_{\mu }k_{\nu }}{(\eta .k)^{2}}\right] 
\end{equation}
{[}that is obtained from the action by inverting the quadratic form in it for
\( k^{2}\neq 0;\; \eta .k\neq 0 \){]} by a modified propagator valid for all
\( k \). The latter, in turn, can be obtained by inverting the quadratic form
in a modified action in \( A \), that formally differs from the original action
by quadratic \( O(\varepsilon A^{2}) \) terms \cite{prep}; where \( \epsilon  \)
is a parameter appearing in the pole prescription. 

In this work,  we wish to elaborate on one of the essential new observation
made in \cite{j02} and to bring out further the power of that observation and
to show that it leads to new conclusions. This observation pertains to the fact
that a careful treatment of the renormalization of gauge theories, formulated
by the path-integral in (\ref{ii}), ought also to take an account the import
of the extra relations that follow from the presence of the residual gauge invariance
as formulated by the IRGT\footnote{%
IRGT stands for the abbreviation of {}``infinitesimal residual gauge transformations{}''
as formulated in \cite{j02}.
} WT-identities in \cite{j02}. These were formulated in \cite{j02} using a
generalized version of infinitesimal residual gauge invariance in the BRS space\footnote{%
These, in particular, deal with the Green's functions with an external momentum
in certain non-trivial domains (such as \( \eta .k=0 \) for axial gauges) and
their form is generally dependent on the specific {}``prescription term{}''.
The content of these \emph{is not} covered by the \emph{usual} BRST WT-identities.
As shown in \cite{j02}, however, the rigorous BRST WT-identity arising from
(\ref{i}), that takes into account the \( \epsilon  \)-term carefully, does
cover IRGT WT-identities.
}. In this work, we wish to draw attention to several observations using these.
We will show, in particular, that a prescription (such as those one considers
commonly \cite{bns,l}), given by any \emph{fixed}\footnote{%
As argued later, the usual ways of giving prescription for poles corresponds
to the addition of a \emph{fixed} term \( \varepsilon O \) in the action.
} \emph{\( \epsilon O \)} term, may lead to the IRGT WT-identities that are
not compatible with the expected form of renormalization together with ghost
decoupling. We shall also show that, if renormalization (in its expected form)
with a given \( \varepsilon O \) is possible\footnote{%
The renormalization scheme with a particular \( \epsilon O \) could, for example,
be obstructed by a lack of validity of usual power counting \cite{bns,l}
} at all, we may generally \emph{need renormalization of the prescription term}
and this possibility, moreover, is not always necessarily consistent with the
IRGT WT-identities. \emph{}Later in this work, we shall formulate the conditions
on \( O \)  under which the latter interpretation becomes possible. As we shall
later see, this observation does not look surprising when seen in the light
of the present framework, where as suggested in \cite{j02}, we may be required
to deal with the \emph{entire action \( S_{eff}+\varepsilon \int d^{4}xO \),
including the symmetry breaking term \( \varepsilon O \)} while discussing
renormalization\emph{.} An obvious question at this point would be why one needs
to care about the renormalization of the \( \epsilon  \)-term at all, if we
are going to take the limit \( \epsilon \rightarrow 0 \) in the answer. This
is suggested by the role of the \( \epsilon  \)-term and the observations made
in \cite{j02} regarding it. In particular, we wish to draw attention to the
fact that the limit \( \epsilon \rightarrow 0 \) in (\ref{ii}) is highly nontrivial
as putting \( \epsilon =0 \) in it leads to an ill-defined path-integral leading
to very many unacceptable consequences \cite{j02}. We will elaborate on it
further at a later stage.

We shall illustrate this point with the help of the axial gauge \( A^{\alpha }_{3}=0 \).
Consider the following set of defining properties and/or assumptions:

\begin{enumerate}
\item \( A_{3}=0 \): spatial axial gauge
\item Multiplicative renormalization of the type:\begin{equation}
\label{mulren}
A_{3}=\widetilde{Z}^{1/2}A^{R}_{3};\; \; \; \; A_{\mu }=Z^{1/2}_{3}A^{R}_{\mu };\; \; \; \mu =0,1,2\; \; \; g=Z_{1}g^{R}
\end{equation}
leading to renormalized Green's functions that are finite and well-defined in
all momentum domains.
\item Ghost decoupling: so that we may assume that the ghost action can be taken as
\begin{equation}
\label{gh}
S_{gh}\equiv \int d^{4}x\{-\overline{c}^{\alpha }\partial _{3}c^{\alpha }+i\varepsilon \overline{c}^{\alpha }c^{\alpha }\}
\end{equation}

\item Path-integral formulation of axial gauges with a prescription for the gauge
propagator poles implemented by a \emph{fixed} quadratic term of the form\footnote{%
We do not include ghosts in O since we have assumed ghost decoupling in 3 above.
} \begin{equation}
\label{pres}
-i\varepsilon \int d^{4}xO[A]=-i\varepsilon \int d^{4}x\int d^{4}yA_{\mu }^{\alpha }(x)a^{\mu \nu }(x,y)A_{\nu }^{\alpha }(y)
\end{equation}
where \( O[A] \) is (generally) a nonlocal operator.
\end{enumerate}
In the following,  we shall first show that the above set is not necessarily
compatible \emph{unless certain additional restrictions, (spelt out later) are
satisfied by \( O[A] \).}

We shall implement the \( A_{3}=0 \) gauge by the use of the Nakanishi-Lautrup
{}``b{}''- field. This method has also been used in the early literature on
axial gauges by Kummer \cite{k75}. We write the generating functional of Green's
functions of the gauge field as \begin{equation}
\label{i}
W[J]=\int DADbDcD\overline{c}\exp \left\{ iS_{eff}+\varepsilon \int d^{4}xO[A]+i\int d^{4}x\; JA\right\} 
\end{equation}
where\[
S_{eff}=S_{0}-\int d^{4}x\; b^{\alpha }A_{3}^{\alpha }+S_{gh}\]
We note that \begin{equation}
\label{gf}
\int Db\exp \left\{ -i\int d^{4}xb^{\alpha }A_{3}^{\alpha }\right\} \sim \prod _{\alpha ,x}\delta (A_{3}^{\alpha }(x))
\end{equation}
has been used in dropping the \( A_{3} \)-dependence in the ghost-action (\ref{gh}).
We may do the same in \( O[A] \) and assume that it has no \( A_{3} \)-dependence.

We now consider the following infinitesimal transformations based on the residual
gauge-invariance of the action (without the \( \epsilon  \)-term) (following
\cite{j02}, we call them the IRGT) with \( \theta ^{\alpha }=\theta ^{\alpha }(x_{0},x_{1},x_{2}) \)\begin{eqnarray}
A_{\mu }^{\alpha }(x)\rightarrow A_{\mu }^{\alpha }(x)+\partial _{\mu }\theta ^{\alpha }-gf^{\alpha \beta \gamma }A_{\mu }^{\beta }(x)\theta ^{\gamma }(x) &  & \nonumber \\
X^{\alpha }(x)\rightarrow X^{\alpha }(x)-\; gf^{\alpha \beta \gamma }X^{\beta }(x)\theta ^{\gamma }(x) &  & \nonumber \\
X\equiv A_{3},b,c,\overline{c,}\, \partial _{3}c & \label{irgt} 
\end{eqnarray}
We note that under this IRGT, \( S_{eff} \) and \( \overline{c}c \) are invariant.
We, now, carry out the IRGT in \( W \) of (\ref{i}) and equate the change
to zero. We thus obtain,\begin{equation}
\label{rwt1}
<<\int d^{4}x\left\{ J_{\mu }^{\alpha }(x)D^{\alpha \gamma }_{\mu }-i\varepsilon \Delta O^{\gamma }[A]\right\} \theta ^{\gamma }(x)>>\; =\; 0
\end{equation}
where we have expressed the change in \( O \) under (\ref{irgt}) as\begin{equation}
\label{do}
\int d^{4}x\; O\rightarrow \int d^{4}x\; O+\int d^{4}x\; \Delta O^{\gamma }\theta ^{\gamma }(x)
\end{equation}
and we have defined, for any X,\begin{equation}
\label{db}
<<X>>\equiv \int DADbDcD\overline{c}X\exp \left\{ iS_{eff}+\varepsilon \int d^{4}xO[A]+i\int d^{4}x\; JA\right\} 
\end{equation}
In view of the fact that \( \theta ^{\gamma }=\theta ^{\gamma }(x_{0},x_{1},x_{2}) \)
can be varied arbitrarily, we find that (\ref{rwt1}) leads us to,\begin{eqnarray}
0\; =\; <<\int dx_{3}\left\{ D^{\alpha \gamma }_{\mu }J^{\gamma \mu }(x)+i\varepsilon \Delta O^{\alpha }[A]\right\} >> &  & \nonumber \\
=\; <<\int dx_{3}\left\{ \sum _{\mu \neq 3}[\partial ^{\mu }J_{\mu }^{\alpha }(x)+gf^{\alpha \beta \gamma }J_{\mu }^{\beta }(x)A^{\gamma \mu }(x)]+i\varepsilon \Delta O^{\alpha }[A]\right\} >> &  & \nonumber \\
=\; \int dx_{3}\left\{ \sum _{\mu \neq 3}\left[ \partial ^{\mu }J_{\mu }^{\alpha }(x)W[J]-igf^{\alpha \beta \gamma }J_{\mu }^{\beta }(x)\frac{\delta W[J]}{\delta J_{\mu }^{\gamma }(x)}\right] +i\varepsilon <<\Delta O^{\alpha }[A]>>\right\}  &  & \label{rwt} 
\end{eqnarray}
In the above, we have dropped the term \( \sim A_{3} \) using the \( \delta - \)function
in (\ref{gf}). We remark that, as emphasized in \cite{j02}, the last term
can have a finite limit as \( \epsilon \rightarrow 0 \) (even in tree approximation)
and its presence cannot just be ignored.

The above identity is over and above the \emph{usual formal} BRST-WT identity
(in which no account of the \( \epsilon  \)-term is taken) and as pointed out
in \cite{j02}, the renormalization has to be compatible (or made compatible)
with it. We now discuss, in the light of (\ref{rwt}), various possibilities
regarding the pole prescription treatment . Before proceeding, we shall note
that

\begin{enumerate}
\item If \( O[A] \) is a local quadratic term \( \sim A_{\mu }^{\alpha }(x)A^{\alpha \mu }(x) \)
then \( \Delta O^{\alpha }[A]\sim \sum _{\mu \neq 3}\partial ^{\mu }A_{\mu }^{\alpha } \)
is linear in A. We further note that under the assumption of the multiplicative
renormalization, \( Z_{3}^{-1/2}\Delta O^{\alpha }[A] \) is a finite operator.
\item If \( \int d^{4}xO[A] \) is a non-local quadratic term \( \int d^{4}x\int d^{4}y\; A_{\mu }^{\alpha }(x)a^{\mu \nu }(x,y)A_{\nu }^{\alpha }(y) \)
then\[
\Delta O^{\alpha }[A]=2\int d^{4}y\left\{ -\partial _{x}^{\mu }a^{\mu \nu }(x,y)A_{\nu }^{\alpha }(y)+gf^{\alpha \beta \gamma }A_{\mu }^{\beta }(x)a^{\mu \nu }(x,y)A_{\nu }^{\gamma }(y)\right\} \]
and has two terms: One is linear in A and the other is quadratic in A and is
moreover a composite operator. We express this, in obvious notations, as \( \Delta O[A]\equiv \Delta _{1}O+\Delta _{2}O \).
\end{enumerate}
\textbf{A SPECTATOR PRESCRIPTION TERM}

It is usually assumed \cite{bns,l}that the prescription for treating the axial
gauge propagator is unaffected by renormalization and so is {}``\( \epsilon  \){}''.
Thus, in this case, we are effectively assuming that the term \( \epsilon O[A] \)
is unaffected during the renormalization process. We shall call this case the
{}`` spectator prescription term{}''.

In case one above of a local quadratic \( O[A] \), the renormalizations of
each of the three terms in (\ref{rwt}) has been assumed to be multiplicative
with scales: Z\( _{3}^{-1/2} \); Z\( _{1} \) and Z\( _{3}^{1/2} \). These
would be compatible only if \( Z_{1}=1=Z_{3} \). This would, of course, contradict
a non-trivial value for \( \beta  \)-function which is (expected to be) gauge-independent
and hence must be the same as the Lorentz gauges.

The discussion for the case 2 above, is a special case of the discussion given
below for the {}`` renormalized prescription term{}'' and we shall see that
it is required that \( O[A] \) must satisfy certain constraints. More comments
are made later.

\textbf{RENORMALIZED PRESCRIPTION TERM}

We shall now explore, however, another (and a more general) possibility in which
the (\ref{rwt}) is made consistent with renormalization. We shall not insist
on keeping the \( \epsilon  \)-term fixed in form, but allow it to be modified
under the renormalization process. Thus we are allowing for a {}``renormalization
of prescription{}''. We shall now explore the restrictions on \( O \), under
which this is possible. We assume that renormalization replaces the \( \varepsilon O[A] \)
term by\footnote{%
With the assumption of ghost-decoupling, \( O[A] \) cannot mix with a \( \overline{c}^{\alpha }c^{\alpha } \)
like operators involving ghosts.
} say \( \varepsilon \{O[A]+\widetilde{O[A]\}} \) (where \( \widetilde{O[A]} \)
depends on the regularization parameter). We need not any further treat \( \epsilon  \)
as a parameter that can be rescaled, as the definition of \( \widetilde{O[A]} \)
can absorb effects of such a scaling. The (\ref{rwt}) then is replaced by the
renormalized version of the (\ref{rwt}), viz.\begin{equation}
\label{renrwt}
\int dx_{3}\left\{ \sum _{\mu \neq 3}\left[ \partial ^{\mu }J_{\mu }^{\alpha }(x)W[J;\varepsilon ]-igf^{\alpha \beta \gamma }J_{\mu }^{\beta }(x)\frac{\delta W[J;\varepsilon ]}{\delta J_{\mu }^{\gamma }(x)}\right] +i\varepsilon <<\Delta O^{\alpha }[A]+\Delta \widetilde{O}^{\alpha }[A]>>\right\} 
\end{equation}
Further analysis of (\ref{renrwt}) will have to be carried out under a restricted
but {}``reasonable{}'' set of assumptions spelt out later in various places.
First of all, we shall assume that \( O[A] \) is of net dimension two. We shall
write, in obvious notations, \( \Delta \widetilde{O[A]}\equiv \Delta _{1}\widetilde{O[A]}+\Delta _{2}\widetilde{O[A]} \);
where the two pieces are respectively linear and quadratic in A\footnote{%
This amounts to the \emph{assumption} that the usual power counting works for
the prescription at hand.
}. We multiply the identity by \( Z^{1/2}_{3} \) and express the equation in
terms of the renormalized quantities\footnote{%
We are going to assume that the renormalized Green's functions are finite functions
of \( \epsilon  \) for \( \epsilon  \) in some interval (0, \( \varepsilon _{0} \)).
We require this especially since in axial gauges, it has been found that there
can be finite contributions to diagrams from \( \varepsilon \bullet \frac{1}{\varepsilon } \)type
terms (See e.g. Ref. \cite{at89}). In any case, the \( \epsilon \rightarrow 0 \)
limit is to be taken only at the end of the calculation.
}:\begin{eqnarray}
\int dx_{3}\sum _{\mu \neq 3}\left[ \partial ^{\mu }J_{\mu }^{R\alpha }(x)W^{R}[J^{R};\varepsilon ]-iZ_{1}Z^{1/2}_{3}g^{R}f^{\alpha \beta \gamma }J_{\mu }^{R\beta }(x)\frac{\delta W^{R}[J^{R};\varepsilon ]}{\delta J_{\mu }^{R\gamma }(x)}\right]  &  & \nonumber \\
=-i\varepsilon Z_{3}\int dx_{3}<<\Delta _{1}O^{\alpha }[A^{R}]+\Delta _{1}\widetilde{O}^{\alpha }[A^{R}]>> &  & \nonumber \label{rwt11} \\
-i\varepsilon Z^{1/2}_{3}\int dx_{3}<<\Delta _{2}O^{\alpha }[A]+\Delta _{2}\widetilde{O}^{\alpha }[A]>> & \label{rwt11} 
\end{eqnarray}
Let us now discuss the above equation in the 1-loop approximation. We express
\( Z_{3}=1+z_{3} \) etc. and look at the divergent part of (\ref{rwt11}).
We find,\begin{eqnarray}
 & i(z_{1}+\frac{1}{2}z_{3})g^{R}f^{\alpha \beta \gamma }J_{\mu }^{R\beta }(x)\frac{\delta W^{R}[J^{R};\varepsilon ]}{\delta J_{\mu }^{R\gamma }(x)} & \nonumber \\
= & i\varepsilon \int dx_{3}<<z_{3}\Delta _{1}O^{\alpha }[A^{R}]+\Delta _{1}\widetilde{O}^{\alpha }[A^{R}]>> & \nonumber \\
 & +i\varepsilon \int dx_{3}<<\Delta _{2}O^{\alpha }[A]>>^{div}_{i}A^{R}_{i}++i\varepsilon \frac{z_{3}}{2}\int dx_{3}\Delta _{2}O^{\alpha }[A^{R}]W^{R}[J^{R}] & \nonumber \\
 & +i\varepsilon \int dx_{3}<<\Delta _{2}O^{\alpha }[A]>>^{div}_{mn}A^{R}_{m}A^{R}_{n}+i\varepsilon \int dx_{3}<<\Delta \widetilde{_{2}O}^{\alpha }[A]>> & \label{rwt12} 
\end{eqnarray}
where we have expressed (in obvious notations) the linear and the quadratic
terms in \( <<\Delta _{2}O[A]>>^{div} \) in one loop approximation\footnote{%
We are again \emph{making an assumption} that the naive power-counting will
work here also. Moreover, note that in evaluating \( <<\Delta _{2}O^{\alpha }[A]>>^{div}_{mn} \)
, we need to pay attention to the fact that there are \emph{unrenormalized}
coupling and fields in \( \Delta _{2}O^{\alpha }[A] \) that do contribute to
the divergence.
}. We note that the usual BRST WT-identities, which hold when one stays away
from external momenta satisfying \( k.\eta =0 \), imply that, should the multiplicative
renormalization as postulated be possible, we have \begin{equation}
\label{z1z3}
z_{1}+\frac{1}{2}z_{3}=0
\end{equation}
 We now compare the \( O[A] \) and \( O[A^{2}] \) terms on both sides :It
leads us to two constraints:\begin{equation}
\label{con1}
0=\int dx_{3}<<z_{3}\Delta _{1}O^{\alpha R}[A]+\Delta _{1}\widetilde{O}^{\alpha R}[A]>>+\int dx_{3}<<\Delta _{2}O^{\alpha }[A]>>^{div}_{i}A^{R}_{i}
\end{equation}
\begin{equation}
\label{con2}
0=\int dx_{3}\left[ \frac{1}{2}z_{3}\Delta _{2}O^{\alpha }[A^{R}]W[0]+<<\Delta _{2}O^{\alpha }[A]>>^{div}_{mn}A^{R}_{m}A^{R}_{n}+<<\Delta _{2}\widetilde{O}^{\alpha }[A^{R}]>>\right] 
\end{equation}
These constraints determine the unknowns \( \Delta _{1}\widetilde{O}^{\alpha }[A^{R}] \)
and \( \Delta _{2}\widetilde{O}^{\alpha }[A^{R}] \). In addition, there is
the requirement that these can be written as the IRGT variation of some \( \int d^{4}x\widetilde{O}[A] \).
Moreover, this term \( \varepsilon \int d^{4}x\widetilde{O}[A] \) when added
to the action should make, say, the inverse propagator \( \Gamma _{\mu \nu }(k,\varepsilon ) \)
in 1-loop finite. If there is a solution to these conditions, then only one
can interpret this as the {}``renormalization of prescription{}''.

To summarize, up-to 1-loop order, the IRGT WT-identity can be made consistent
with renormalization in the assumed form by {}`` renormalization of prescription{}''
if :

{[}1{]} There exists an \( \int d^{4}x\widetilde{O}[A] \), such that its IRGT
variation \( \Delta \widetilde{O} \) can be expressed as \( \Delta \widetilde{O[A]}\equiv \Delta _{1}\widetilde{O[A]}+\Delta _{2}\widetilde{O[A]} \)
; where \( \Delta _{1}\widetilde{O}^{\alpha }[A^{R}] \) and \( \Delta _{2}\widetilde{O}^{\alpha }[A^{R}] \)
satisfy the constraints (\ref{con1}) and (\ref{con2}) ; 

{[}2{]} The counterterm \( \varepsilon \int d^{4}x\widetilde{O}[A] \) makes
\( \Gamma _{\mu \nu }(k,\varepsilon ) \) finite;

{[}3{]} The usual power counting holds to this order for the renormalization
of \emph{nonlocal} operator \( \Delta _{2}O^{\alpha }[A] \).

(These spell out the sufficient conditions).

Finally, we note that the case 2 of a {}``spectator prescription term{}''
is a special case of the above discussion with \( \Delta \widetilde{O} \) deleted.
Thus, in this case, it is necessary that (\ref{con1}) and (\ref{con2}) hold
with the terms \( \widetilde{\Delta _{1}O} \) and \( \widetilde{\Delta _{2}O} \)
deleted. 

We add some conclusions that follow from an analysis of the above conditions.
The analysis of these conditions shows that:

\begin{enumerate}
\item Let us suppose that the (arbitrary) function \( a_{\mu \nu }(x-y)=a_{\nu \mu }(y-x) \)
in (\ref{pres}) be such that the power counting in momentum space \emph{in
terms of the external momenta} holds for the one-loop diagrams contributing
to \( <<\Delta _{2}O^{\alpha }[A]>>_{i} \) and \( <<\Delta _{2}O^{\alpha }[A]>>_{mn} \)
in the sense that the divergence in the first is a monomial in \( p \) of degree
1; and that in latter a monomial of degree zero. {[}Note: \( \int d^{4}x\Delta _{2}O^{\alpha }[A]\equiv 0 \){]}.
\item Then, in momentum space, \( <<\Delta _{2}O^{\alpha }[A]>>^{div}_{i} \) is of
the form \( p_{\mu }\Delta ^{\mu \nu } \) (with \( \Delta ^{\mu \nu } \) a
\emph{constant} matrix); and the divergence \emph{from the one-loop diagram}
contributing to \( <<\Delta _{2}O^{\alpha }[A]>>^{div}_{mn} \) vanishes on
account of the fact that \( \int d^{4}x\Delta _{2}O^{\alpha }[A]\equiv 0 \)
for any \( a_{\mu \nu } \).
\item In such a case, a solution to the conditions (\ref{con1}) and (\ref{con2})
exists provided \( \Delta ^{\mu \nu } \)is symmentric for (\( \mu  \),\( \nu  \)=
0,1,2) and is given by,\begin{equation}
\label{sol}
\int d^{4}x\widetilde{O[A]}=-z_{3}\int d^{4}xO[A]+\frac{1}{2}\int d^{4}x\Delta '^{\mu \nu }A^{\alpha }_{\mu }A^{\alpha \mu }
\end{equation}
where \( \Delta '_{\mu \nu }=\Delta _{\mu \nu } \) for all \( (\mu ,\nu ) \)
except that \( \Delta '_{3i}=\Delta _{i3};i=0,1,2 \) .
\item If we further assume that the divergence in \( \Gamma _{\mu \nu }(p,\varepsilon ) \)
proportional to \( \epsilon  \) is , by the assumed validity of power counting
in terms of external momenta, also a constant, and therefore independent of
momentum \( p \), then the condition {[}2{]} above is also satisfied by this
solution.
\end{enumerate}
The above solution (\ref{sol}) has been given under certain conditions sufficient
for its existence. The main restriction on \( O \) seems to come from (1) the
requirement of power-counting as enumerated above; and (2) the symmetry requirement
on \( \Delta ^{\mu \nu } \) mentioned above in 2.

We shall note further that while the present analysis has arrived at its results
using a specific form of path-integral definition of non-covariant gauges of
(\ref{i}), we expect an equivalent set of conclusions should follow from any
other way of defining these gauges. This formalism has enabled us to see the
existence of and to arrive at these conclusions in a easy and direct manner.
No such analysis seems to have been carried out in the context of attempts at
defining the non-covariant gauges \cite{bns,l} in other ways.

\textbf{A QUALITATIVE EXPLANATION}

We shall now explain the results qualitatively. Consider the inverse propagator
\( \Gamma _{\mu \nu } \) for the gauge-field in one loop approximation. There
is a contribution to the \( \epsilon  \)-dependent terms to this order. For
momenta \( k \), such that \( k.\eta \neq 0 \), the \( \epsilon  \) terms
as a whole are negligible (as \( \epsilon \rightarrow 0 \)). In this sector,
the usual multiplicative renormalization does the job of making the inverse
propagator finite, if \( \epsilon  \)-terms are ignored. Nonetheless, in the
\emph{3-dimensional} subspace \( k.\eta =0 \), the quantity \( k^{\mu }\Gamma _{\mu \nu }k^{\nu } \)
obtained by taking the longitudinal projection of \( \Gamma  \) has only \( \epsilon  \)-terms
remaining \emph{}(and the inverse of which tends to infinity as \( \epsilon \rightarrow 0 \)\emph{).
These also receive divergences ; which need not generally be removed by the
field-renormalization.} (Recall that there was no such subspace in the case
of Lorentz gauge that needs to be worried about). One may be required to perform
an extra renormalization on the \( \epsilon  \)-term (This may have to be checked
in each case).

At this point, one may ask the justifiable question, as to whether the renormalization
of the \( \epsilon  \)-term should matter at all, since we mean to take the
\( \epsilon  \) to zero in the end!. Earlier, we have already made some comments
based on \cite{j02}. In addition, we recall that there are several examples
\cite{bns,l} where the change of prescription has altered (1) the nature and
the presence of divergences (2) value of gauge-invariant quantities\footnote{%
Here, we recall that two different prescriptions \( \epsilon O \) and \( \epsilon O' \)
\emph{may not be related by a residual gauge transformation,} and hence they
need not lead to identical physical results. Moreover, neither of these need
coincide with the Lorentz gauge result for analogous reasons.
}. This makes us strongly suspect that this sector in momentum space is important
enough.

Now, \( S_{eff} \) is invariant under IRGT. Any prescription breaks the residual
gauge invariance in a particular manner. It is not obvious that the physical
quantities so calculated using it will be gauge-independent. This is controlled
by the behavior of the path-integral under infinitesimal residual gauge transformations
as formulated by IRGT WT-identities\footnote{%
As mentioned earlier, these have been shown to be contained in the BRST-identities
for the \emph{net} action including the \( \epsilon  \)-term in \cite{j02}.
}. Under IRGT, the path-integral changes solely due to the {}``symmetry breaking{}''
term \( \epsilon O \) in addition to the source term. The form of divergence
in the variation in the source term is restricted by the \emph{assumptions}
we made in the beginning. This restriction then becomes imposed on the divergences
that can arise from the variation of the \( \epsilon  \)-term via IRGT WT-identity
(and such terms can have non-vanishing contributions as \( \epsilon \rightarrow 0 \)
\cite{j02}). These are additional restrictions on \( O \), and it not \emph{a
priori} obvious that they will be obeyed. 

\textbf{SUMMARY AND CONCLUSIONS}

We shall now summarize our conclusions. We considered the formalism for non-covariant
gauges presented in \cite{j02}, where the {}``prescription{}'' is imposed
via an \( \varepsilon \int d^{4}xO[A] \) term added to the action. We found
this formalism lead us in an easy manner to an additional consideration that
is required in the definition and renormalization of these gauges. We illustrated
this for the \( A_{3}=0 \) gauge. This fact, which was brought out in \cite{j02},
has been further elaborated and analyzed here. We see that the usual expectations
of multiplicative renormalization together with ghost decoupling are not automatically
compatible with every prescription term \( \varepsilon \int d^{4}xO[A] \);
there are additional constraints that have to be satisfied further by it (which
are implied by the IRGT WT-identities). We also pointed out the need to have
to deal with renormalization of \( \epsilon  \)-terms carefully. These considerations
do not seem to have been taken into account so far in attempts to define noncovariant
gauges. 

\textbf{ACKNOWLEDGMENT}

I would like to acknowledge support from Department of Science and Technology,
India via grant for the project No. DST/PHY/19990170.


\begin{thebibliography}{10}
\bibitem{bns}A. Bassetto, G. Nardelli, and R. Soldati, Yang-Mills Theories in Algebraic Non-covariant
Gauges (World Scientific, Singapore, 1991) and references therein.
\bibitem{l}G. Leibbrandt, Non-covariant Gauges (World Scientific, Singapore, 1994) and
references therein. See also, Physical and non-standard gauges (Springer Verlag 1990)P.Gaigg, W.Kummer and M.Schweda (Editors)
\bibitem{1}See  references 5-9 for some recent works; also see the references therein.
\bibitem{dev}See e.g. references in {[}1,2{]} as well as ref. {[} 8,9{]} and those therein.
\bibitem{jm00}S. D. Joglekar, and A. Misra, Int. J. Mod. Phys.A 15, 1453 (2000); Erratum \emph{ibid}A15,
3899(2000); S. D. Joglekar, and A. Misra, J. Math. Phys. 41, 1755-1767 (2000).
\bibitem{jm99}S. D. Joglekar, and A. Misra, Mod. Phys. Lett. A14, 2083 (1999); \textbf{\emph{ibid}}
A15, 541-546 (2000); Int. J. Mod. Phys.A 16, 3731 (2001); S. D. Joglekar, Mod.
Phys. Lett. A15, 245-252 (2000); Int. J. Mod. Phys.A 16, 5043 (2001); S. D.
Joglekar, and B. P. Mandal, Int. J. Mod. Phys.A 17, 1279 (2002).
\bibitem{j02}S.D.Joglekar {}``Some Observations on Non-covariant Gauges and the \( \epsilon  \)-term{}''-
hep-th/0205045 ;Mod.Phys. Lett.A 17, 2581 (2002)
\bibitem{bz}L. Baulieu, and D. Zwanziger, Nucl. Phys. B548, 527-562 (1999) .
\bibitem{leib}G. Leibbrandt, Nucl. Phys. Proc. Suppl. 90, 19 (2000); G. Heinrich and G. Leibbrandt,
Nucl. Phys B575,359 (2000)
\bibitem{k75}W. Kummer, Nucl. Phys. B 100, 106 (1976)
\bibitem{prep}S.D.Joglekar (in preparation).
\bibitem{at89}A. Andrasi and J.C.Taylor, Nucl.Phys. \textbf{B310},222 (1988)\end{thebibliography}
\end{document}